\documentclass[conference]{IEEEtran}

\usepackage{cite}
\usepackage{amsmath,amssymb,amsfonts}
\usepackage{amsthm}
\usepackage{graphicx}

\usepackage{algorithm}
\usepackage{setspace} 
\usepackage{algpseudocode} 
\usepackage{cite}
\usepackage{hyperref}
\usepackage{subfigure}

\usepackage [font=scriptsize]{caption}
\usepackage{url}

\def\BibTeX{{\rm B\kern-.05em{\sc i\kern-.025em b}\kern-.08em
    T\kern-.1667em\lower.7ex\hbox{E}\kern-.125emX}}

\theoremstyle{definition}

\begin{document}
%
\title{An Efficient Scheme for the Generation of Ordered Trees in Constant Amortized Time}


\author{{Victor Parque, Tomoyuki Miyashita}\\
\IEEEauthorblockA{Department of Modern Mechanical Engineering, \\ Waseda University, \\
Shinjuku 3-4-1, Tokyo, 169-8555, Japan\\parque@aoni.waseda.jp}
}

\maketitle

\begin{abstract}
Trees are useful entities allowing to model data structures and hierarchical relationships in networked decision systems ubiquitously. An ordered tree is a rooted tree where the order of the subtrees (children) of a node is significant. In combinatorial optimization, generating ordered trees is relevant to evaluate candidate combinatorial objects. In this paper, we present an algebraic scheme to generate ordered trees with $n$ vertices with utmost efficiency; whereby our approach uses $\mathcal{O}(n)$ space and $\mathcal{O}(1)$ time in average per tree. Our computational studies have shown the feasibility and efficiency to generate ordered trees in constant time in average, in about one tenth of a millisecond per ordered tree. Due to the 1-1 bijective nature to other combinatorial classes, our approach is favorable to study the generation of binary trees with $n$ external nodes, trees with $n$ nodes, legal sequences of $n$ pairs of parentheses, triangulated $n$-gons, gambler's sequences and lattice paths. We believe our scheme may find its use in devising algorithms for planning and combinatorial optimization involving Catalan numbers.
\end{abstract}



%
\IEEEpeerreviewmaketitle

\begin{IEEEkeywords}
ordered trees, plane trees, catalan trees, enumeration, combinatorial objects, encoding, graphs, constant amortized time, Catalan numbers, lattice paths, algorithms
\end{IEEEkeywords}

\section{Introduction}

Trees often arise as fundamental mechanisms to model data structures and hierarchical relationships in networked decision systems ubiquitously. For instance, trees enable to model the optimal connectivity among disjoint entities in an environment\cite{aldana20}, allow to model the efficient routing in distribution systems\cite{marques20,he20}, and allow to study candidate plans for information fusion\cite{zhou16,zhangconcrete20} and assembly\cite{watson19} over classes of combinatorial problems in decision making.

Well-known books on graphs and combinatorial objects have studied trees in considerable scrutiny\cite{goldberg93,kreher98,wilf89}. Basically, four major types of trees arise ubiquitously: (1) the free tree, which is an acyclic connected graph, (2) the rooted tree, which is a free tree with a distinguished root node, (3) the ordered tree, which is a rooted tree where the order of the subtrees (children) of a node is significant, and (4) the binary tree, which is an ordered tree where every node has degree 0 or 2\cite{sefla}.

In this paper, we study the problem of generating ordered trees arbitrarily and exhaustively. In the literature, an ordered tree is also well-known as plane tree and Catalan tree. The term plane is due to trees being able to be transformed from one topology to another topology by using continuous operations in the plane. Also, due to the Catalan number representation, ordered trees have 1-1 correspondence to binary trees with $n$ external nodes, trees with $n$ nodes, legal sequences of $n$ pairs of parentheses, triangulated $n$-gons, gambler's ruin sequences, and lattice paths.

The problem of generating ordered trees and their related structures has received favorable attention. For instance, Pallo proposed the use of integer sequences under a tailored order to generate ordered trees with $n$ vertices and $k$ leaves in $O(n-k)$ time\cite{pallo87}, Nakano presented the algorithms to generate rooted ordered trees with at most $n$ vertices in $O(n)$ space and $O(1)$ time per tree in average\cite{naka02}. Also, extensions were proposed to generate rooted ordered trees with exactly $n$ vertices in $O(1)$ time in average, and rooted ordered trees with exactly $n$ vertices and $k$ leaves in $O(n-k)$ time. The basic approach in \cite{naka02} is based on reversing the removal of the rightmost path of rooted trees so that the entire genealogy of rooted ordered trees having at most $n$ edges is traversed.

In line of the above, the idea of using traversing the genealogy (or family tree) was later extended to generate ordered trees with specified diameter in $O(1)$ time per tree \cite{naka05}, to generate bipartite permutation graphs in $O(1)$ in the worst case\cite{saito12}, and to generate floor plans subject to a rectangle and $n$ points in the plane. Also, Yamanaka et al. presented the enumeration of ordered trees with exactly $n$ vertices and $k$ leaves in $O(1)$ time in the worst case\cite{yaka09}. Generally speaking, the above approaches are based on the reverse search method\cite{reverse96} which basically consists in generating objects through a graph (tree) whose edges model local and bounded operations on the objects, thus the exhaustive generation of objects (trees) is possible by traversing the graph backwards by an adjacency expansion oracle.

By using the main generation principle of Ruskey and Hu\cite{ruskeyhu77}, Beyer and Hedetniemi used a level sequence representation to generate rooted ordered trees with $n$ vertices in reverse lexicographic order. Their approach achieved Constant Amortized Time (CAT) overall trees, that is in $O(1)$ time per tree in average\cite{beyer80}. This method has been reported in the book of Wilf\cite{wilf89} as well. Beyer and Hedetniemi were the first to introduce the Constant Amortized Time (CAT) property. Li and Ruskey used a \emph{parent array} representation and recursive algorithms based on depth first traversal to generate rooted and free trees\cite{liruskey99}. Also, the more natural representation of parent arrays allowed to model constraints in height and parenthood seamlessly. Sawada used level sequence representation to generate all circular ordered trees in $O(1)$ time per tree on average\cite{sawa06}. Knuth presented the approach to generate all free trees with $O(n)$ space and $O(1)$ time per tree in average\cite{knuth06}. Sen-Peng Eu et al. presented the enumerations and bijections of vertices of rooted ordered trees with constraints in levels and degrees\cite{sen17}.

Atkinson presented the linear time algorithm to generating binary trees at random\cite{at92}. Knott presented bijection algorithms between binary trees of $n$ vertices and a segment of integers\cite{knott77}, Rotem and Varol used a ballot sequence representation of stack-sortable permutations to allow bijections between binary trees and these combinatorial objects\cite{rotem78}, and Solomon and Finkel modified Knott's algorithm by using recursions and binary searches to allow ranking in $O(n)$, unranking in $O(n\log n)$ and generating the sucessor of a given tree in $O(n)$ \cite{solo80}. Jing and Tang presented the generation of all binary trees by using the triangulation of convex polygons \cite{jing16}. In \cite{zer85}, Zerling presented a tailored codeword whose sum lies in $[0, n-1]$ and whose bijection to rooted binary trees is feasible in $O(n)$.

\begin{figure*}[ht!]
	\centering
	\includegraphics[width=0.78\textwidth]{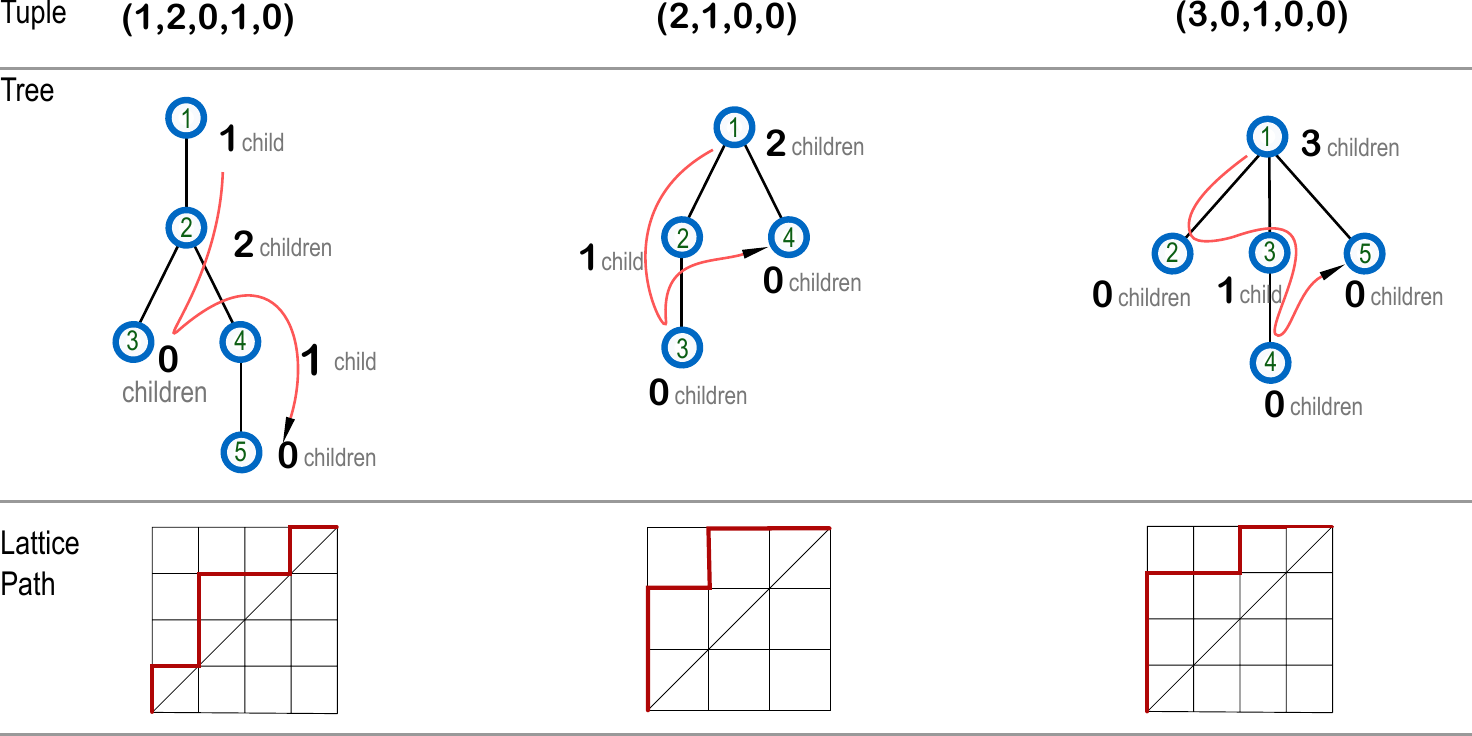}
	\caption{Examples of tree encoding showing the number of children of each node in preorder. To represent the lattice path, the digits in the tuple encode the relative column height in which the first $n - 1$ digits are used, the last zero digit of the tuple is ignored.}
	\label{enco}
\end{figure*}

Although the above-mentioned studies enable to generate ordered trees arbitrarily and exhaustively with utmost efficiency, the existing approaches are unable to be easily extrapolated to combinatorial optimization problems involving ordered trees. One of the main reasons is that the above approaches are often based on operations on the tree structures; however, in combinatorial optimization settings, it would be desirable to generate trees based on numbers portraying lower and upper bounds on the search space, and sample from there based on an oracle or heuristic. In this paper, we present an algebraic approach to generate ordered trees. In the best of our knowledge, our proposed approach is the first proposing algebraic and straightforward operations to generate ordered trees. In particular, our contributions are summarized as follows:

\begin{itemize}
\item an algebraic approach to generate the arbitrary and the complete set of ordered trees with $n$ nodes. Our approach uses $\mathcal{O}(n)$ space and $\mathcal{O}(1)$ time in average per tree.
\item the computational studies showing the feasibility and efficiency to generate ordered trees in constant time in average, in about one tenth of a millisecond per ordered tree.
\end{itemize}

Due to the bijective nature to other combinatorial classes, we believe our approach is also useful to generate binary trees with $n$ external nodes, trees with $n$ nodes, legal sequences of $n$ pairs of parentheses, triangulated $n$-gons, gambler's ruin sequences, lattice paths and other combinatorial objects based on catalan numbers.

In the rest of this article, we describe our approach and discuss our computational experiments.

\section{Proposed Method}

In this section, we present the main concepts involved in our proposed approach.

\subsection{Encoding Mechanism}

In an ordered tree, the children of every node is ordered, that is, there exists a first child, second child, third child, and so on. In order to represent an ordered tree, we use an $n$-tuple to encode a tree with $n$ nodes. Here,

\begin{equation}\label{t}
t = (t_1, t_2, ..., t_i, ...,  t_n)
\end{equation}
, denotes a tree, in which $t_i$ represents the number of children of the $i$-th node of the tree $t$. Thus, a parent node implies $t_i >0$, a leaf (terminal) node implies $t_i = 0$.

The above-mentioned representation is inspired on the BCT representation of binary trees\cite{eeckman94}, in which B denotes branching, C denotes continuation and T denotes a terminal. Basically, the BCT representation of a tree can be rendered from the sum of entries below the diagonal of the adjacency matrix and, viceversa, the adjacency matrix is renderable from the BCT encoding by an $O(n)$ algorithm based on stacks\cite{cuntz10d}. Thus, assuming nodes are pre-labeled by a user-defined order, the elements of the tuple $t$ are equivalent to the sum of the rows of the adjacency matrix representation of the tree.

In order to exemplify the encoding we use here, Fig. \ref{enco} shows examples of tree encoding for various tree topologies and number of nodes; also Fig. \ref{enco} shows the equivalence of the tree encoding (tuples) to generate lattice paths. In Fig. \ref{enco}, the reader may note that node labels are defined by the order of traversal. For simplicity and without loss of generality, we traverse a tree in preorder (by visiting first the root, then by visiting the subtrees from left to right). Also, due the above encoding requiring the traversal from root to children and then to leaf (terminal) nodes, a natural consequence is that the last element of the tuple $t$ is a terminal node; thus for a tree with $n$ nodes,
\begin{equation}\label{condt}
t_n = 0, ~ t_i \in [0, n-1].
\end{equation}

Furthermore, due to the tuple $t$ encoding the number of children in its elements, the following holds

\begin{equation}\label{sumbct}
\sum_{i = 1}^{n} t_i = n-1
\end{equation}
, since the $t_i$ is equivalent to the number of branches (edges) at a given node. Thus, the sum of all elements of the tuple $t$ is equivalent to the number of edges of the tree.

\subsection{Arbitrary Ordered Trees}

Here, by using the above-mentioned encoding, we propose a mechanism to generate arbitrary ordered trees. It is possible to generate ordered trees by finding each element $t_i$ of the tuple $t$ for $i \in [n]$ by

\begin{equation}\label{tisample}
t_i \sim  \mathbf{U}\{L_i, U_i\} ~ i = 1, 2, ..., n
\end{equation}

\begin{equation}\label{Li}
L_i = 1 - \text{sgn} \Big (  t_{i-1} + S_{i-1} - 1 \Big )
\end{equation}

\begin{equation}\label{Ui}
U_i = U_{i-1} - t_{i-1}
\end{equation}

where $L_i$ and $U_i$ are the lower and upper bound on $t_i$, respectively, such that $t_i \in [L_i .. U_i]$, and sgn(.) denotes the signum function. Since the above mechanism is recursive in nature, the initial conditions $S_0 = 0$, $U_0 = n$, $t_0 = 1$ are necessary. The variable $S$ is used to compute the accumulation of elements of the tuple. It is possible to eliminate the variable $S$, by which an equivalent expression to Eq. \ref{Ui} can be obtained

\begin{equation}\label{Uiv2}
L_i = 1 - \text{sgn} \Bigg (  \sum_{j = 0}^{i-1} (t_j - 1) \Bigg )
\end{equation}

For ordered trees with $n$ nodes, and considering the ordered nature of the tuple $t$ and of Eq. \ref{Li} and Eq. \ref{Ui}, the following relations hold $L_1 = 1$, $U_1 = n-1$, and $L_n = U_n = 0$, thus $t_n = 0$, which aligns well with Eq. \ref{condt}, and

\begin{equation}\label{t1}
t_1 \in [1..n-1].
\end{equation}

\subsection{Genealogy of Ordered Trees}

Let $T_n$ be the set of ordered trees with $n$ nodes. The size of $T_n$ can be computed from the binomial coefficient

\begin{equation}\label{Tn}
| T_n | = \frac{1}{n}\binom{2(n-1)}{n-1}.
\end{equation}

As can be seen from Eq. \ref{Tn}, $T_1 = 1$, $T_2 = 1$, $T_3 = 2$, $T_4 = 5$, $T_5 = 14$, $T_6 = 42$, ... represent the well-known Catalan numbers.

\begin{figure}[t!]
	\centering
	\includegraphics[width=0.4\columnwidth]{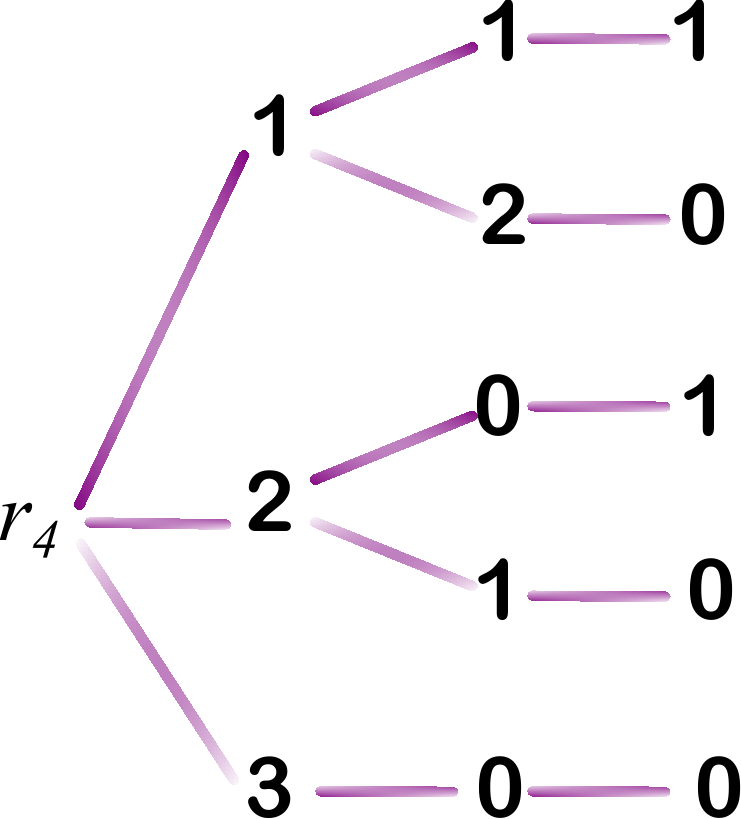}
	\caption{Genealogical tree $T_4$.}
	\label{family4}
\end{figure}

\begin{figure}[t!]
	\centering
	\includegraphics[width=0.5\columnwidth]{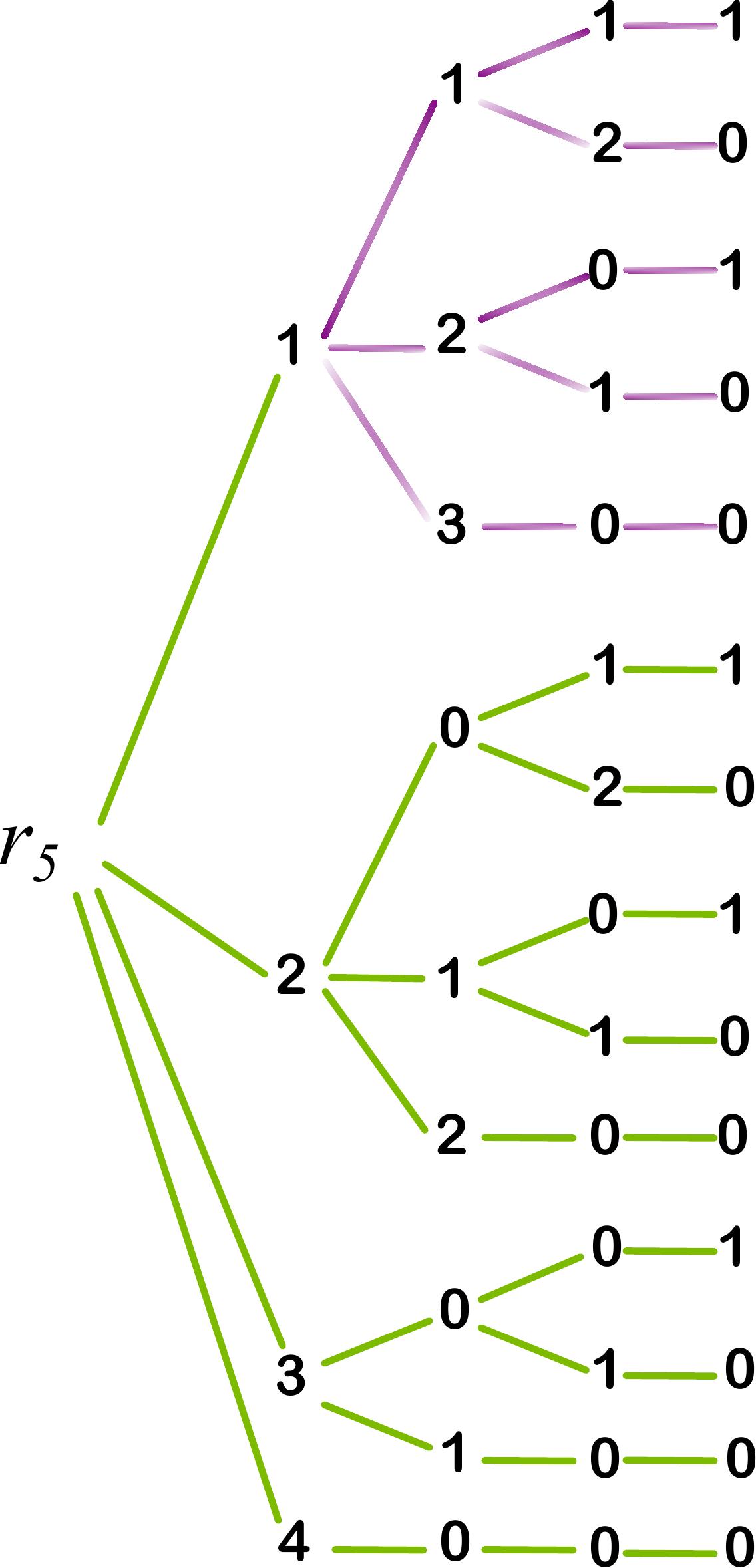}
	\caption{Genealogical tree $T_5$.}
	\label{family5}
\end{figure}

Since the bounds for $t_i$ are given as

\begin{equation}\label{tibound}
t_i \in [L_i .. U_i],
\end{equation}
it becomes possible to generate all ordered trees with $n$ nodes exhaustively by generating each integer $t_i$ from $[L_i .. U_i]$ consecutively for $i = 1, 2, ..., n-1$. Here, we skip the case $i = n$ due to the fact of Eq. \ref{condt}. Thus, $T_n$ can regarded as a genealogical tree rooted at $r_n$ with internal nodes being labeled with numbers in the interval $[0 .. n-1]$. To exemplify our argument, Fig. \ref{family4} and Fig. \ref{family5} show the genealogical trees $T_n$ for $n = 4$ and $n= 5$. The reader may note that the number of leaves coincide with the size of $T_n$, that is the Catalan numbers expressed by Eq. \ref{Tn}, and that $T_{n-1}$ is a substructure of $T_{n}$.

By traversing all nodes from the root $r_n$ to each leaf of $T_n$, it is possible to obtain integer sequences of the form

\begin{equation}\label{seq}
r_n: ~(t_1, t_2, ..., t_{n-1}, 0),
\end{equation}
which is in line with the representation expressed by Eq. \ref{t}. The term "0" is tacit due to Eq. \ref{condt}.

Let $T^i_n$ be the $i$-th tree of the set $T_n$. Then, generating ordered trees by sampling integer numbers from the interval $[L_i .. U_i]$ in ascending order implies that the first tree is $(1, 1, 1, ..., 0)$ and the last tree is $(n-1, 0 ,0 , ..., 0)$. Conversely, if one uses the descending order instead, the opposite occurs, that is the first tree is $(n-1, 0 ,0 , ..., 0)$ and the last tree is $(1, 1, 1, ..., 0)$. The above implies that we can generate all ordered trees without repetition, which leads to achieve $\mathcal{O}(1)$ time in average to generate each tree.

Furthermore, generating all ordered trees with $n$ nodes implies that we can generate all the set $T_n$ by traversing all leaves of $T_n$ by a recursive approach, as portrayed by Algorithm \ref{gen}. As such, ordered trees will be generated by a sequence of numbers within the same line. Algorithm \ref{gen} shows numbers in same lines (new lines) to imply a forward (backward) traversal from the root (leaves) towards the proximal root (leaves). Assuming we use the ascending order approach to generate numbers $[L_i .. U_i]$ in Fig. \ref{family4}, the first, second and third ordered trees are encoded by $(1,1,1)$, $(1,2,0)$, and $(2,0,1)$ respectively (the 0 at the end is omitted for the sake of simplicity). Thus Algorithm \ref{gen} will show $1 ~1~ 1$ in the first line, $2~ 0$ in the second line, and $2~ 0~ 1$ (the spaces between numbers are added for the sake of clarity). The above implies that to obtain the second tree, it will be necessary to transform $1 ~1~ 1$ to $(1,2,0)$ by changing its last two digits. Thus, for an ordered tree with $n$ nodes, the maximum number of changes needed to generate a new ordered tree is $n-1$. As such, Algorithm \ref{gen} does not output entire trees, instead it outputs the difference from the previous ordered tree. Due to the above-mentioned considerations, our approach uses $\mathcal{O}(n)$ space. Storing all the structure of $T_n$ is unnecessary.

\begin{algorithm}[t!]
\caption{Generate Genealogy}\label{gen}
\begin{spacing}{1.38}
\begin{algorithmic}[1]
\Procedure{Genealogy}{$L, U, s, n$}

\If{$n > 0$ }

\For{$i \gets L \text{ \textbf{to} } U$}
\State $u \gets U - i$
\State $s' \gets s + i - 1$
\State $l \gets 1 - \text{sgn}(s')$
\State Show digit $i$
\State \Call{Genealogy}{$l, u, s', n-1$}
\State Show a New Line

\EndFor\label{defor}
\EndIf
\EndProcedure
\end{algorithmic}
\end{spacing}
\end{algorithm}

\begin{figure*}[ht!]
	\begin{center}
		\subfigure[Total time for all trees]{\includegraphics[width=0.48\textwidth]{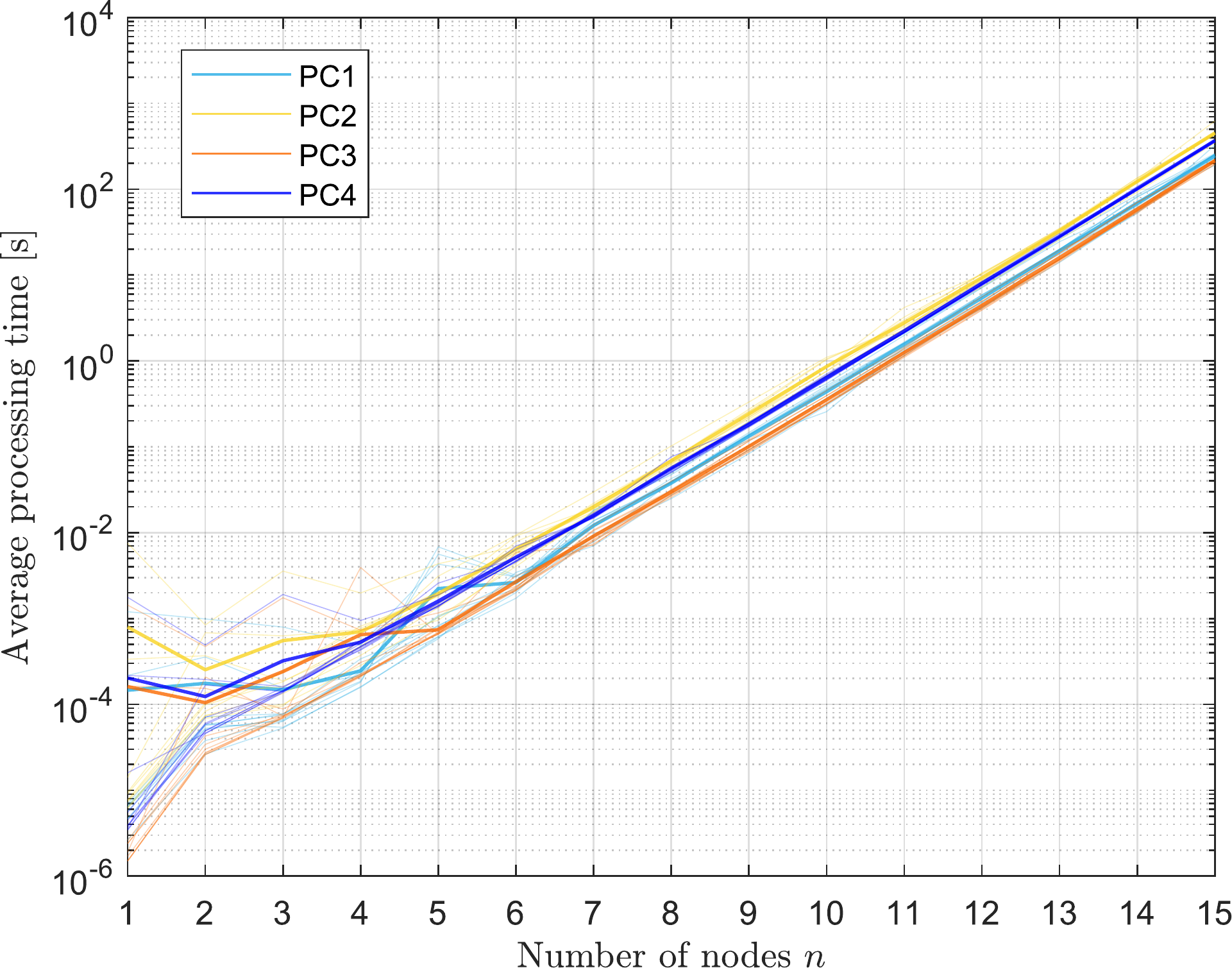}}
		\hfill
		\subfigure[Average time per tree]{\includegraphics[width=0.48\textwidth]{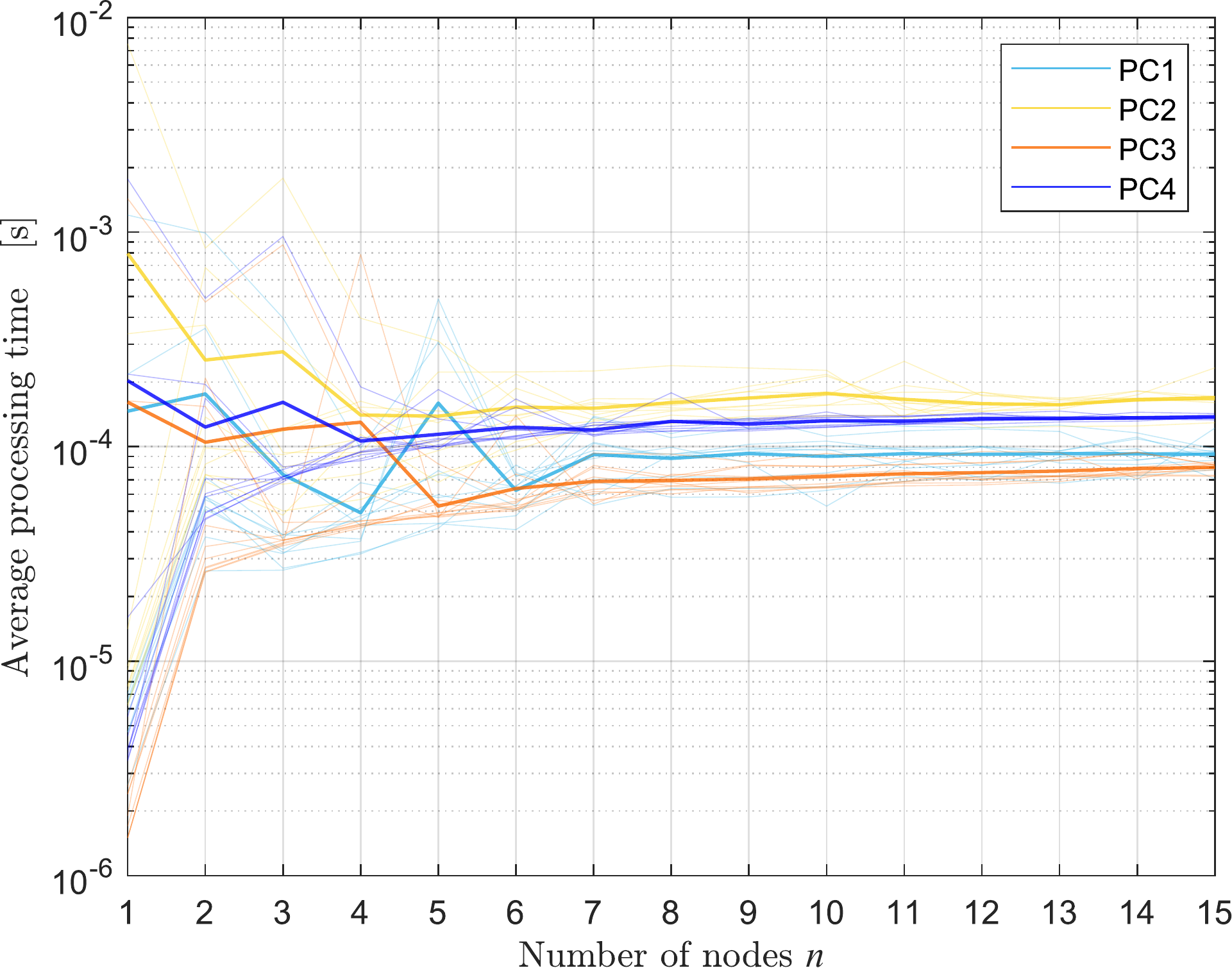}}
		\hfill
	\end{center}
	\caption{Time to generate all ordered trees.}
	\label{time}
\end{figure*}

\section{Computational Experiments}

In order to evaluate the feasibility and efficiency of our proposed approach, we performed computational experiments comprising the generation of ordered trees with distinct number of nodes. Our approach was implemented in Matlab 2020a and evaluations occurred considering:

\begin{itemize}
  \item Number of nodes $n = 1, 2, ..., 15$.
  \item For each value of $n$, all trees were generated over 10 independent runs.
  \item The time to generate all trees for each $n$ and each independent run was logged.
  \item To compare the performance in distinct hardware configurations, we used 4 types of computing environments.
\end{itemize}

By following the above-mentioned configurations, and due the sizes of $T_n$, expressed as Eq. \ref{Tn} for $n = 1, 2, ..., 15$, we generated 148,314,080 trees. The computing environments used were as follows:

\begin{itemize}
  \item PC1. Intel Core i7 @2.8GHz (4 CPUs), 16 GB RAM
  \item PC2. Intel Core i7 @2.9GHz (4 CPUs), 16 GB RAM
  \item PC3. AMD Ryzen Threadripper 2990WX @ 3.0 GHz (64 CPUs), 128 GB RAM
  \item PC4. Intel Core i7 @3.4GHz (12 CPUs), 64 GB RAM
\end{itemize}

In order to show the time performance and efficiency frontiers involved in our algorithm, Fig. \ref{time} shows (a) the total time used to generate ordered trees and (b) the average time per tree as a function of the number of nodes. For the sake of clarity both plots show the log-scale comparisons. As we can see from Fig. \ref{time}-(a), the time needed to generate all ordered trees show a linear-like behaviour. We believe this is due to relatively smaller values of $n$ up to 15. In theory, Fig. \ref{time}-(a) is expected to behave in the order $\mathcal{O}(|T_n|)$ due to the Catalan numbers expressed Eq. \ref{Tn}. Investigating the experimental landscape for very large $n$ is left to future work.

By observing Fig. \ref{time}-(b), the time required to generate each tree is constant in average, within a tenth of a millisecond. We believe this is due to the simplicity and straightforward algebraic approach used to generate ordered trees. Here, the performance over independent runs are shown by thinner lines, and the mean over independent runs is show by a thicker line. Also, by observing Fig. \ref{time}-(b), the performance across distinct computing environments is similar between one another, wherein the relatively improved performance is lead by clock speed, which is in line with the straightforward algebraic computations.

\begin{figure}[ht!]
	\centering
	\includegraphics[width=0.88\columnwidth]{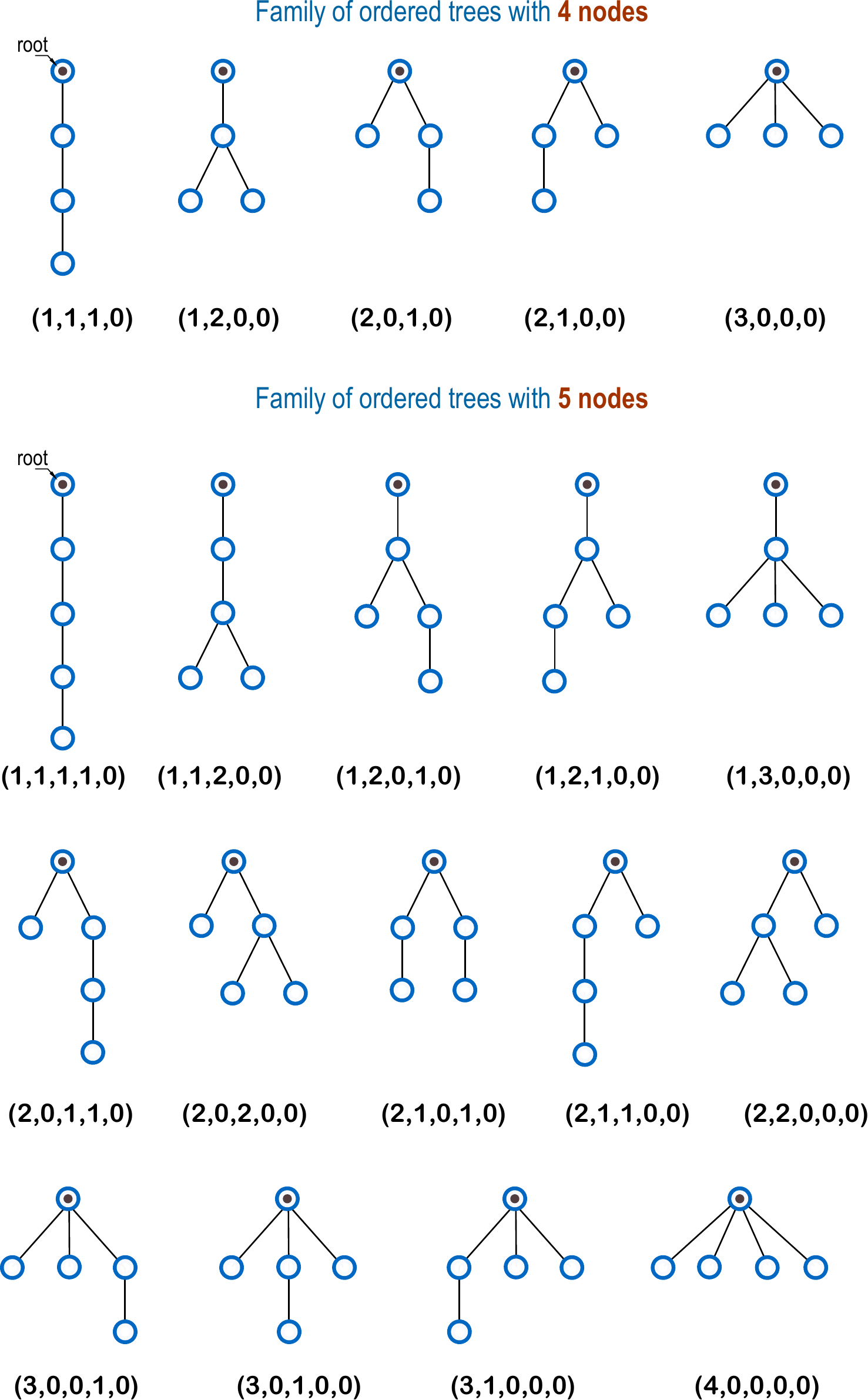}
	\caption{Ordered trees with $n = 4$ and $n=5$ nodes.}
	\label{trees45}
\end{figure}

\begin{figure}[ht!]
	\centering
	\includegraphics[width=0.98\columnwidth]{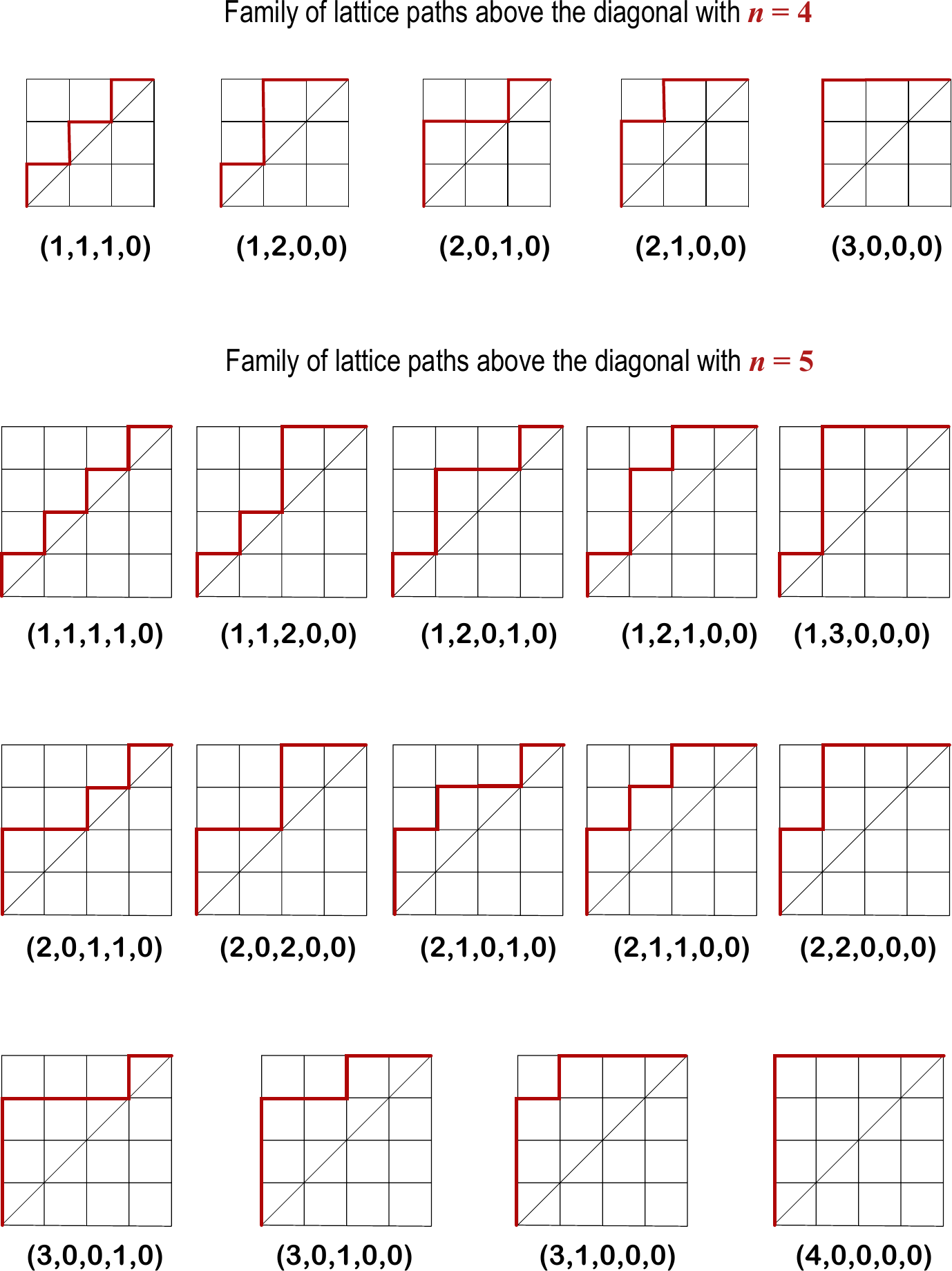}
	\caption{The 1-1 correspondence with the lattice paths above the diagonal of a grid with $(n-1) \times (n-1)$ square cells. Top: the case of $n = 4$. Bottom: the case of $n=5$. To represent the lattice paths, the digits encode the relative column height in which only the first $n-1$ digits are used, the last digit (0) is ignored.}
	\label{lattice45}
\end{figure}

In order to show the kind of trees that our approach is able to generate, Fig. \ref{trees45} shows the encoding and topologies of ordered trees for $n=4$ (top) and $n= 5$ (bottom). Here, ordered trees are shown in the order of top-down of Fig. \ref{family4} and Fig. \ref{family5}, that is by traversing all leafs of $T_n$ by the recursive Algorithm \ref{gen}, and following $t_i \in [L_i.. U_i]$ in ascending order. Note that according to the observation in Eq. \ref{condt}, the encoding of trees involve the last element of every encoding to be $t_n = 0$ for all cases. As we can observe from Fig. \ref{trees45}, our approach is able to generate the complete set of trees. Due to 1-1 correspondence with other combinatorial objects involving Catalan numbers, our approach can be used to generate lattice paths above the diagonal, as Fig. \ref{lattice45} shows. Here, lattice paths use the relative column height in a grid of $(n-1)\times (n-1)$ cells.

The above observations bring implications to generate binary trees with $n$ external nodes, trees with $n$ nodes, legal sequences of $n$ pairs of parentheses, triangulated $n$-gons, gambler's ruin sequences, lattice paths and other combinatorial objects based on catalan numbers. In future work, we aim at studying the performance for very large $n$ and its applications in combinatorial optimization in Robotics and Operations Research, e.g. reactive motion planning\cite{henrich98,gecks09} and binomial graphs\cite{bingraphs17}. We believe the proposed encoding and generation mechanism may find its use in planning and combinatorial optimization involving catalan numbers.

\section{Final Notes}

In this paper, we have proposed an algebraic approach to generate the arbitrary and the complete set of ordered trees with $n$ nodes. Our approach uses $\mathcal{O}(n)$ space and $\mathcal{O}(1)$ time in average per tree. By using computational studies, we have also shown the feasibility and efficiency to generate ordered trees in constant time in average, which in practice translates in about one tenth of a millisecond per ordered tree. Due to the 1-1 correspondence to other combinatorial classes, our approach is also useful to generate binary trees with $n$ external nodes, trees with $n$ nodes, legal sequences of $n$ pairs of parentheses, triangulated $n$-gons, gambler's ruin sequences, lattice paths and further combinatorial objects based on catalan numbers. Further studies in our agenda aim at studying the performance frontiers for very large $n$ and its applications in combinatorial optimization in Robotics and Operations Research.

%

\bibliographystyle{IEEEtran}
\bibliography{mybiblio}



%

\end{document}